\documentstyle[12pt]{article}
\newcommand{\bc}{\begin{center}}
\newcommand{\be}{\begin{equation}}
\newcommand{\bra}[1]{\left.\langle#1\right|}
\newcommand{\bea}{\begin{eqnarray}}

\newcommand{\ec}{\end{center}}
\newcommand{\ee}{\end{equation}\noindent}
\newcommand{\eea}{\end{eqnarray}}
\newcommand{\eps}{\varepsilon}
\newcommand{\gapp}{\stackrel{>}{\sim}}

\newcommand{\ket}[1]{\left|#1\rangle\right.}

\newcommand{\rslash} [1]{\mbox{#1\hspace{-7pt}/}}

\newcommand{\SS}{\scriptstyle}

\newcommand{\vect}[1]{\mbox{\boldmath$#1$}}
\newcommand{\bdot}{\mbox{\boldmath$\cdot$}}

\begin{document}
\begin{center} \large
\bf{On Size and Shape of the Average Meson Fields\\
in the Semibosonized Nambu \& Jona-Lasinio Model}
\normalsize\end{center}
\vskip\baselineskip \vskip\baselineskip
\begin{center}
{\sc R.~W\"unsch$^1$, K.~Goeke$^2$, Th.~Mei{\ss}ner$^3$}
\end{center}
\bigskip \bigskip

$^1$ Institut f\"ur Kern- und Hadronenphysik, Forschungszentrum
Rossendorf e.V.,\\ \hspace*{9mm} Postfach 51\,01\,19, 01\,314 Dresden,
Germany\\

$^2$ Institut f\"ur Theoretische Physik II, Ruhr-Universit\"at Bochum,\\
\hspace*{9mm} Postfach 10\,21\,48,\ 44\,780\ Bochum, Germany\\

$^3$ Institute for Nuclear Theory, University of Washington, HN-12
Seattle,\\  \hspace*{9mm} WA 98\,195, USA

\vskip\baselineskip
\vskip\baselineskip
\vskip\baselineskip

{\bf{Abstract:}}
We consider a two-flavor Nambu \& Jona-Lasinio model in Hartree
approximation involving scalar-isoscalar and pseudoscalar-isovector
quark-quark interactions.  Average meson fields are defined by minimizing
the effective Euklidean action. The  fermionic part of the action, which
contains the full Dirac sea, is regularized within Schwinger's proper-time
scheme.  The meson fields are restricted to the chiral circle and to
hedgehog configurations. The only parameter of the model is the constituent
quark mass $M$ which simultaneously controls the regularization.

We evaluate meson and quark fields self-consistently in dependence on the
constituent quark mass. It is shown  that the self-consistent fields do
practically not depend on the constituent quark mass.
This allows us to define a properly parameterized reference field which for
physically relevant constituent masses can be used as a good approximation
to the exactly calculated one. The reference field is chosen to have
correct behaviour for small and large radii.

To test the agreement between self-consistent and reference fields we
calculate several observables like nucleon energy, mean square radius,
axial-vector constant and delta-nucleon mass splitting in dependence on
the constituent quark mass.  The agreement is found to be very well.\\[4mm]

\newpage

\section{Introduction}

The model of Nambu \& Jona-Lasinio (NJL) \cite{Nam-61} has
been used quite successfully as effective chiral theory for low and medium
energy hadronic phenomena. First it has been applied to vacuum and meson
properties as well as medium effects (for reviews
c.~f.~\cite{Vog-91,Kle-92,Hat-94}). Later on it turned out that also
baryonic systems (nucleons and hyperons) can be described within this model
(for a review c.~f.~\cite{Mei-94}).
Starting from a semi-bosonized version \cite{Egu-76,Kle-76} with
scalar-isoscalar and pseudoscalar-isovector interaction and treating the
meson fields classically various authors have shown that for constituent
quark masses $M\gapp350\,MeV$ it is possible to get self-consistent
solitonic solutions with baryon number B=1 consisting of 3 valence quarks
in addition to the polarized Dirac sea \cite{Dya-88}-\cite{Wak-91}.
Because of the non-renormalizability of the Nambu \& Jona-Lasinio model
the sea contribution diverges and has to be regularized. The parameters of
the model can be fixed to the physics of the meson and vacuum
sector, mainly to the weak pion decay constant $f_\pi$ and the pion mass
$m_\pi$. In doing so only one free parameter remains open for the baryonic
sector for which we take the constituent quark mass $M$.

The self-consistent determination of the meson fields is a time-consuming
numerical procedure. Changing the parameters of the model or the
regularization scheme the procedure has to be repeated again.
So it might be helpful to look for an analytic parametrization of the
selfconsistent profile function $\Theta(r)$ of the solitonic solution which
approximates the exact $\Theta(r)$ as well as possible. Within the
restrictions to hedgehog configurations \cite{Cho-75} and to the chiral
circle the meson fields are uniquely described by the profile function
$\Theta(r)$.  In the course of our calculations we noticed to a very large
extent an independence of this profile function on the constituent quark
mass $M$.  It is the aim of this paper to investigate this dependence
quantitatively and to look for a general function which may approximate the
profile function, if possible independently of $M$.

In section 2 we review the main ideas of the semi-bosonized and
regularized Nambu \& Jona-Lasinio model for two flavors and introduce
observables characterizing the quark and meson configuration.
The dependendence of the self-consistent meson profiles $\Theta(r)$ on the
constituent quark mass $M$ is investigated in sect.~3 within a wide range
($350\,MeV\le M\le1000\,MeV$). Here we compare the self-consistent profile
with a reference profile $\Theta^{Ref}(R;r)$ given by a simple
arithmetic expression obtained from an asymptotic expansion of the equation
of motion at $r\to0$ and $r\to\infty$.  In sect.~4 we evaluate several
observables like nucleon mass, mean-square radius, axial-vector coupling
constant and delta-nucleon mass splitting using both the  self-consistently
determined profiles and the standard profiles. The comparison of both
values tests and illustrates the quality of the reference profile.

\section{The regularized and bosonized Nambu \& Jona-Lasinio model and
its observables}

The details of the following section can be found in
ref.~\cite{Mei-94,Mei-91a,Wak-91,Mei-91}. Here we shortly review
these parts of the formalism which make this paper self-contained.

We consider a two-flavor NJL lagrangian

\be\label{1}
{\cal{L}}_{NJL}\left(\bar{q}q\right)\, =\,
\bar{q}[i\rslash{$\partial$}-m]q +
\frac{G}{2}\left[(\bar{q}q)^2+(\bar{q}i\gamma_5\hat{\vect{\tau}}q)^2\right]
\end{equation}
for the quark fields $q(x)$ ($u$ and $d$ quarks of $N_c=3$ colours).
Here $G$ is the coupling constant, $\hat{\vect{\tau}}$ the vector of
Pauli-matrices and $m$ the average current mass of the light quarks
$m=(m_u+m_d)/2$.
The theory with only quark degrees of freedom is converted into an
effective quark-meson theory by means of standard path-integral
bosonization \cite{Egu-76,Kle-76,Ebe-86}. The mesonic fields are restricted
to hedgehog configurations and to the ciral circle. The resulting
semi-bosonized theory is described by an effective (Euklidean) action,
whose dynamical quantity is the profile function $\Theta(r)$
\cite{Rei-88,Mei-89}

\be\label{2}
{\cal{A}}_{eff}\left[\Theta\right]\,=\,
{\cal{A}}^q\left[\Theta\right]+{\cal{A}}^m\left[\Theta\right].
\ee
The ${\cal{A}}_{eff}[\Theta]$ consists of a quark part

\be\label{3}
{\cal{A}}^q\left[\Theta\right]\,=\,
-Sp\,ln\left[\beta\left(\frac{\partial}{\partial\tau}+h-\mu\right)\right]+
Sp\,ln\left[\beta\left(\frac{\partial}{\partial\tau}+h_V\right)\right]
\ee
with the quark hamiltonian

\be\label{4}
h\,=\,\vect{\alpha}\bdot\vect{p}+gf_\pi\beta\left[cos\Theta(r)
+i\,sin\Theta(r)\,\gamma_5\,\hat{\vect{\tau}}\bdot\hat{\vect{r}}\right],
\ee
and of a meson part

\be\label{5}
{\cal{A}}^m\left[\Theta\right]\,=\,
{\cal{T}}mf_\pi\frac{\lambda^2}{g}\,4\pi\!\int\!r^2dr
\left[1-cos\Theta(r)\right],
\ee
where ${\cal{T}}$ is the total Euklidean time interval.
The quark-meson coupling strength $g$ in the hamiltonian (\ref{4}) and the
parameter $\lambda$ in the mesonic action (\ref{5}) are related to the
interaction strength $G$ in the NJL lagrangian (\ref{1})  via
$G=g^2/\lambda^2$.
Since we are interested in stationary field configurations we
have assumed the meson fields to be time-independent. Hence the eigenvalues
$\eps_\alpha$ of the hamiltonian (\ref{4}) are real. The vaccum state
(with broken chiral symmetry) has $\Theta(r)\equiv0$ and
and the eigenvalues and eigenfunctions of corresponding
hamiltonian are denoted by $\eps_\alpha^V$ and $\Phi_\alpha^V(\vect{r})$,
respectively.  The symbol $Sp$ indicates functional and matrix (spin,
isospin, colour) trace

\be\label{6}
Sp\,{\cal{O}}\,\equiv\,
N_c\,tr_\gamma\,tr_\tau\int\!d^4x_E\bra{x_E}{\cal{O}}\ket{x_E}
\ee
with the Euklidean space-time vector $x^\mu_E=(\tau,\vect{r})$ and its
volume element $d^4x_E$.
The effective quark action (\ref{3}) can be divided into a valence
contribution

\be\label{7}
{\cal{A}}^q_{val}[\Theta]\,\equiv\, {\cal{A}}^q[\Theta](\mu)-
{\cal{A}}^q[\Theta](\mu=0)\,=\,
-{\cal{T}}\frac{N_c}{2}\sum_\alpha
\left[\left|\eps_\alpha-\mu\right|-\left|\eps_\alpha\right|\right]
\ee
and a sea contribution

\be\label{8}
{\cal{A}}^q_{sea}[\Theta]\,\equiv\, {\cal{A}}^q[\Theta](\mu=0)\,=\,
-{\cal{T}}\frac{N_c}{2}\sum_\alpha
\left[\left|\eps_\alpha\right|-\left|\eps^V_\alpha\right|\right].
\ee
The valence contribution (\ref{7}) depends on the chemical potential
$\mu$ which will be adjusted such that the resulting soliton has baryon
number one. This can be achieved by a chemical potential $\mu>0$ which is
slightly larger than the energy $\eps_{val}$ of the lowest quark level
with positive energy (valence level). However, the meson field may be so
strong that the valence level is bent down to the Dirac sea
($\eps_{val}<0$). In this case the Dirac sea has one level more than in
vacuum state and therefore carries baryon number one.  Describing a baryon
with baryon number one we have to choose $\mu=0$, in this case, making the
Fermi sea ($0\le\eps_\alpha\le\mu$) empty.

The sea contribution (\ref{8}) diverges and is regularized. Applying
Schwinger's proper-time scheme $ln\,{\cal{O}}\,\longrightarrow\,
\left.ln\,{\cal{O}}\right|_{Reg}\,=\, -\int_{1/\Lambda^2}^{\infty}
\frac{ds}{s} e^{-s{\cal{O}}}\:$ \cite{Sch-51}, where $\Lambda$ is the
regularization parameter, we get a regularized sea contribution

\be\label{9}
{\cal{A}}_{sea}^{q,Reg}\,=\,
-{\cal{T}}\frac{N_c}{2}\sum\limits_\alpha
\left[R_E(\eps_\alpha,\Lambda)|\eps_\alpha|-
R_E(\eps_\alpha^V,\Lambda)|\eps_\alpha^V|\right].
\ee
Here the regularization function is given by

\be\label{10}
R_E(\eps,\Lambda)\,=\,-\frac{1}{\sqrt{4\pi}|\eps|}
\int\limits_{1/\Lambda^2}^\infty dt\,t^{-3/2}e^{-\eps^2t}\,=\,
-\frac{1}{\sqrt{4\pi}}
\:\Gamma\!\left(-\frac{1}{2},\frac{\eps^2}{\Lambda^2}\right)
\ee
with the incomplete Gammafunction $\Gamma(x,a)$.  The total regularized
effective action reads

\be\label{11}
{\cal{A}}_{eff}^{Reg}[\Theta]\,=\, {\cal{A}}^q_{val}[\Theta]+
{\cal{A}}^{q,Reg}_{sea}[\Theta]+ {\cal{A}}^m[\Theta].
\ee
The profile function $\Theta(r)$ is determined by

\be\label{12}
\frac{\delta{\cal{A}}_{eff}^{Reg}[\Theta]}{\delta\Theta(r)}\,=\,0.
\ee
yielding the equation of motion

\be\label{13}
\Theta(r)\,=\,arc\,tan\frac{\bar{P}(r)}{\bar{S}(r)}
\ee
with

\be\label{14}
\bar{S}(r)\,=\,\frac{m}{g}-\frac{g}{\lambda^2}
\left[N_c\,\theta(\eps_{val})\,\bar{S}_{val}(r)
-\frac{N_c}{2}\sum\limits_\alpha
R_m(\eps_\alpha,\Lambda)\,\bar{S}_\alpha (r)\right]
\ee
and

\be\label{15}
\bar{S}_\alpha(r)\,=\,\frac{1}{4\pi}
\int\!d\hat{\vect{r}}\,\bar{\Phi}_\alpha(\vect{r})\,\Phi_\alpha(\vect{r}),
\ee

\be\label{16}
\bar{P}(r)\,=\,-\frac{g}{\lambda^2}
\left[N_c\,\theta(\eps_{val})\,\bar{P}_{val}(r)
-\frac{N_c}{2}\sum\limits_\alpha
R_m(\eps_\alpha,\Lambda)\,\bar{P}_\alpha (r)\right]
\ee
and

\be\label{17}
\bar{P}_\alpha(r)\,=\,\frac{1}{4\pi}
\int\!d\hat{\vect{r}}\,\bar{\Phi}_\alpha(\vect{r})\,
i\gamma_5\hat{\vect{r}}\bdot\hat{\vect{\tau}}\,\Phi_\alpha(\vect{r}).
\ee
The valence contributions $\bar{S}_{val}(r)$ and $\bar{P}_{val}(r)$ are
obtained for the valence level $\alpha=val$. They vanish if the valence
level dips into the Dirac sea ($\eps_{val}<0$). The $\Phi_\alpha(\vect{r})$
are the eigenfunctions of $h$.  The sea contributions have been regularized
within the proper-time scheme with a regularization function

\be\label{18}
R_m(\eps,\Lambda)\,=\,\frac{\eps}{\sqrt{\pi}}
\int\limits_{1/\Lambda^2}^\infty dt\,t^{-1/2}e^{-\eps^2t}\,=\,
\frac{sign(\eps)}{\sqrt{\pi}}
\:\Gamma\!\left(\frac{1}{2},\frac{\eps^2}{\Lambda^2}\right).
\ee

The parameters of the model will be fixed by adjusting the vaccuum and
meson sectors as described in detail in ref.~\cite{Mei-91a}. The only
remaining free parameter is the constituent quark mass

\be\label{19}
M\,=\,gf_\pi
\ee
which will be varied within reasonable limits.

In this paper we consider the (isoscalar) baryon density

\be\label{20}
\rho(\vect{r})\,=\,
\frac{1}{N_c}\left\langle:
q^\dagger(\vect{r})\,q(\vect{r}):\right\rangle\,=\,
\rho(\vect{r})_{val}+\rho(\vect{r}\,)_{sea} \ee with

\be\label{21}
\rho(\vect{r})_{val}\,=\,
N_c\,\theta(\eps_{val})\,\Phi_{val}^\dagger(\vect{r})\,\Phi_{val}(\vect{r})
\ee
and

\be\label{22}
\rho(\vect{r})_{sea}\,=\,
-\frac{1}{2}\sum\limits_\alpha\left[
sign\left(\eps_\alpha\right)\,
\Phi_\alpha^\dagger(\vect{r})\,\Phi_\alpha(\vect{r})-
sign\left(\eps_\alpha^V\right)\,
\Phi_\alpha^{V\dagger}(\vect{r})\,\Phi_\alpha^V(\vect{r})\right].
\ee
The colon in the matix element of eq.~(\ref{20}) indicates subtraction
of the corresponding value calculated in the vacuum state. Like for any
other observable the valence contribution (\ref{21}) vanishes if the Fermi
sea is empty. The Dirac sea contribution (\ref{22}) has not been
regularized because it is finite and fulfills the exact normalization
condition (baryon number zero/one) only in the unregularized case
\cite{Mei-91a}.  The size of the density distribution is characterized by
the mean square baryon radius

\be\label{23}
\bar{R}\,\equiv\,\sqrt{\left<
R^2\right>}\,=\, \left[\int\!d^3r\,r^2\,\rho(\vect{r})\right]^{1/2}.
\ee
Another quantity characterizing a quark configuration is  the
axial density

\be\label{24}
A_o(\vect{r})\,=\,\left\langle:q^\dagger(\vect{r})
\frac{\sigma_o\tau_o}{2}q(\vect{r}):\right\rangle.
\ee
It consists of a valence contribution

\be\label{25}
A_o(\vect{r})_{val}\,=\,
N_c\,\theta(\eps_{val})\,
\Phi^\dagger_{val}(\vect{r})\,\frac{\sigma_o\tau_o}{2}\,\Phi_{val}(\vect{r})
\ee
and of a sea contribution

\bea\label{26}
A_o(\vect{r})_{sea}\,=\,
-\frac{N_c}{2}\sum\limits_\alpha\left[R_m(\eps_\alpha,\Lambda)
\Phi^\dagger_\alpha(\vect{r})\,
\frac{\sigma_o\tau_o}{2}\,\Phi_\alpha(\vect{r})-\right.
\left.R_m(\eps_\alpha^V,\Lambda)
\Phi^{V\,\dagger}_\alpha(\vect{r})\,
\frac{\sigma_o\tau_o}{2}\,\Phi^V_\alpha(\vect{r})\right],
\eea
which has been regularized with the regularization functions
$R_m(\eps,\Lambda)$ defined in eq.~(\ref{18}) \cite{Mei-91}.
The axial density determines the axial-vector coupling constant of the
proton

\be\label{27}
g_A\,=\,-2\int\!d^3r\,A_o(\vect{r}),
\ee
where an additional factor ($-1/3$) is incorporated which results from the
projection onto the isospin quantum number $T=1/2$ of the proton
\cite{Adk-83}.

The total energy $E$ of a static quark-meson configuration is given by
\cite{Mei-91a}

\be\label{28}
E\,=E_{val}^q+E_{sea}^{q,Reg}+E^M+E^{CSB}
\ee\\[0mm]
with the valence-quark energy

\bea\label{29}
E_{val}^q&=&\frac{1}{\cal{T}}
\left[1-\mu\frac{\partial}{\partial\mu}\right]{\cal{A}}_{val}^q\,=\,
\frac{N_c}{2}\sum\limits_\alpha\left[
sign\left(\mu-\eps_\alpha\right)+sign\left(\eps_\alpha\right)
\right]\,\eps_\alpha\nonumber\\[2mm]
&=&N_c\,\theta(\eps_{val})\,\eps_{val}
\eea
and regularized sea-quark contributions

\be\label{30}
E_{sea}^{q,Reg}\,=\,
-\frac{N_c}{2}\sum\limits_\alpha\left[
R_E(\eps_\alpha,\Lambda)\left|\eps_\alpha\right|-
R_E(\eps_\alpha^V,\Lambda)\left|\eps_\alpha^V\right|\,\right].
\ee
The meson energy is given by

\be\label{31}
E^m\,=\,
mf_\pi\frac{\lambda^2}{g}\,4\pi\!\int\!r^2dr
\left[1-cos\Theta(r)\right].
\ee
For time-independent fields the classical meson and sea-quark energies
differ from the corresponding effective actions (\ref{5}, \ref{9}) only by
a factor ${\cal{T}}$.  An additional contribution to the energy results
from $\mu$ dependence of the valence-quark action (\ref{7}).

Finally we consider two spurious contributions to the quark energy which
result from the mean-field approximation and have to be
subtracted from the total energy (\ref{28}). As shown in
\cite{Mei-94,Pob-92} the static hedgehog contains a center-of-mass motion
with the energy

\be\label{32}
E_{CMM}\,=\,\frac{\left\langle:\vect{P}^2:\right\rangle}{2E}\,=\,
\frac{\left\langle:\int\!d^3r\,q^\dagger(\vect{r})\,
(-{\vect{\SS\nabla}}^2)\,q(\vect{r}):\right\rangle}{2E},
\ee
where E is the rest mass represented by the total hedgehog energy
(\ref{28}).  The expectation value of the square
of the total quark momentum $\vect{P}$ consists of a valence- and a
sea-contribution given by

\be\label{33}
\left\langle\vect{P}^2\right\rangle_{val}\,=\,N_c\,\theta(\eps_{val})
\int\!d^3r\Phi_{val}^\dagger(\vect{r})\,
(-\vect{\SS\nabla}^2)\,\Phi_{val}(\vect{r})
\ee
and

\bea\label{34}
\left\langle\vect{P}^2\right\rangle_{sea}\,=\,-\frac{N_c}{2}\sum\limits_\alpha
\left[R_m(\eps_\alpha,\Lambda)
\int\!d^3r\Phi_\alpha^\dagger(\vect{r})\,
(-\vect{\SS\nabla}^2)\,\Phi_\alpha(\vect{r})-\right.\nonumber\\
\hspace*{60mm}\left.R_m(\eps_\alpha^V,\Lambda)
\int\!d^3r\Phi_\alpha^{V\,\dagger}(\vect{r})\,
(-\vect{\SS\nabla}^2)\,\Phi_\alpha^V(\vect{r})\right].
\eea
Another correction term results from the quark rotational degrees of
freedom \cite{Adk-83,Coh-86,Rei-89}. It is described by a moment of
inertia $I$, which can be calculated within the semiclassical cranking
approach \cite{Rin-80} and consists of a valence contribution

\be\label{35}
I_{val}\,=\,\frac{N_c}{2}\,\theta(\eps_{val})
\sum\limits_{\beta\ne val}
\frac{\bra{\Phi_val}\tau_3\ket{\Phi_\beta}\,
\bra{\Phi_\beta}\tau_3\ket{\Phi_val}}{\eps_\beta-\eps_\alpha}
\ee
and of a regularized sea contribution

\be\label{36}
I_{sea}\,=\,\frac{N_c}{2}
\sum\limits_{\alpha\beta}R_I(\eps_\alpha,\eps_\beta;\Lambda)\,
\frac{\bra{\Phi_\alpha}\tau_3\ket{\Phi_\beta}\,
\bra{\Phi_\beta}\tau_3\ket{\Phi_\alpha}}{\eps_\beta-\eps_\alpha}.
\ee
Within the proper-time scheme the regularization function is given by

\bea\label{37}
\lefteqn{R_I(\eps_\alpha,\eps_\beta;\Lambda)\,=}\\[2mm]
&=&\frac{1}{2}\frac{1}{\sqrt{4\pi}}
\int_{1/\Lambda^2}^\infty\!ds\,s^{-\frac{3}{2}}
\frac{1}{\eps_\beta+\eps_\alpha}
\left[e^{-s\eps_\alpha^2}-e^{-s\eps_\beta^2}+
s\,(\eps_\beta-\eps_\alpha)
\left(\eps_\alpha e^{-s\eps_\alpha^2}+\eps_\beta e^{-s\eps_\beta^2}\right)
\right]\nonumber\\[2mm]
&=&\frac{1}{4}
\left[sign(\eps_\beta)\,erfc\left(\frac{|\eps_\beta|}{\Lambda}\right)-
sign(\eps_\alpha)\,erfc\left(\frac{|\eps_\alpha|}{\Lambda}\right)-
\frac{2}{\sqrt{\pi}}
\frac{e^{-\left(\frac{\eps_\alpha}{\Lambda}\right)^2}
     -e^{-\left(\frac{\eps_\beta}{\Lambda}\right)^2}}
{\left(\eps_\alpha+\eps_\beta\right)/\Lambda}\right].\nonumber
\eea
In the limit $\Lambda\to\infty$ one gets the well-known
Inglis formula \cite{Ing-54} for the moment of inertia. The incomplete
error-function is given by
$erfc(x)=\frac{2}{\sqrt{\pi}}\int_x^\infty\!e^{-t^2}\,dt$.
Since the energy correction for the nucleon and the $\Delta$ isobar are
different the moment of inertia gives rise to a mass splitting between
both particles which is given by \cite{Rei-89}

\be\label{38}
E_{\Delta}-E_{N}\,=\,\frac{3}{2I}.
\ee

\section{The self-consistent meson profile and its dependence on the
constituent quark mass}

The equation of motion (\ref{13}) is an implicite and nonlocal equation for
the profile function $\Theta(r)$. Because of the dependence of the
expectation values $\bar{S}(r)$ and $\bar{P}(r)$ on the eigenvalues
and eigenfunctions of the hamiltonian $h$ the right-hand side of the
equation of motion is a functional of the profile function.  We determine
$\Theta(r)$ iteratively. Starting from a reasonable profile $\Theta^o(r)$
we determine eigenfunctions $\Phi_\alpha^o(\vect{r})$ and eigenvalues
$\eps_\alpha^o$ by diagonalizing the hamiltonian (\ref{4}) within an
appropriate basis. By means of the equation of motion (\ref{13}) and the
auxileary functions (\ref{14}-\ref{17}) we get an improved profile
function $\Theta^1(r)$. Continuing this procedure to convergency we get the
self-consistent profile function.

We represent the eigenfunctions of the hamiltonian within a discrete
basis introduced in \cite{Kah-84}. It is defined within a spherical box
with a radius $D$ which is several times larger than the extension of
$\Theta$ field. Details of the procedure are described in
\cite{Rei-88,Mei-89,Mei-91a}.

For all our calculation a box radius $D=15/M$ has turned out to be
sufficiently large. The discrete basis was limited by a maximal momentum
$K_{max}=8\,M$, which is more than 4 times larger than the corresponding
regularization parameter $\Lambda$.

We have numerically determined the self-consistent profiles for 13 values
of the constituent quark mass $M$ between 350 and 1000\,MeV.
As illustrated in fig.~1 the
profiles are nearly independent of the mass parameter. None of the
calculated profiles leaves the narrow corridor limited be the two broken
lines. At masses lower than 350\,MeV the iteration converges to the vaccum
field ($\Theta(r)\equiv0$). Here the corresponding interaction strength $G$
is not strong enough to keep a solitonic configuration together.

The self-consistent profiles can be approximated by a
reference profile

\be\label{39}
\Theta^{Ref}(R;r)\,=\,\left\{ \begin{array}{lll}
-\pi\left(1-\frac{r}{2R}\right)&\quad\mbox{if}&r\le R_M \\[3mm]
-\pi\left(1-\frac{R_M}{2R}\right)\left(\frac{R_M}{r}\right)^2
\frac{1+m_\pi r}{1+m_\pi R_M}\,e^{-m_\pi\left(r-R_M\right)} &
\quad\mbox{if}&r\ge R_M \end{array}\right.\hspace*{1cm}
\ee
with the matching point\\

\be\label{40}
R_M\,=\,\frac{4}{3}R\left(1+\frac{8}{27}m_\pi^2R^2+{\cal{O}}[(m_\pi
R)^3]\right) \hspace*{2cm} \ee
and an empirically determined radius parameter

\be\label{41}
R=0.42\,fm.  \ee
The reference profile (\ref{39}) with the matching point (\ref{40})
interpolates smoothly between the correct asymptotic behaviour at $r\to0$
and $r\to\infty$ following from the asymptotic expansion of the equation of
motion (\ref{13}).
It constitutes - at least visually - a fair
approximation of the self-consistent profiles obtained after a
time-consuming iteration procedure. Further tests of the reference profile
will be performed in the next section.

\section{Testing the reference profile on nucleon observables}

In sect.~2 we considered several expectation values
characterizing a quark or meson configuration. All these quantities are
functionals of the profile function within our model. We have tested the
quality of the reference profile (\ref{39}) and evaluated the observables
using both the reference and the self-consistently determined profiles.
Fig.~2 shows the total energy (\ref{28}) and its components
(\ref{29}, \ref{30}, \ref{31}), the mean square radii ({\ref{23}) of the
baryon density, including their valence- and sea-quark contributions, and
the axial-vector coupling constant of the proton (\ref{27}) calculated with
either profile.  The kinks in the valence and sea contributions at the
critical constituent mass $M=M_{dip}\approx750\,MeV$ result from the
valence level which leaves the Fermi sea and joins the Dirac sea.  The
behaviour of the regularization functions (\ref{10}, \ref{18}, \ref{37}) at
$\eps\to 0$ guarantees that the sum of valence and regularized sea
contributions is a smooth function of the constituent quark mass $M$.
Fig.~2 illustrates that nicely.

The only noticeable difference between the values for self-consistent and
reference profiles appears in the valence and sea contributions in the
vicinity of $M_{dip}$. For the self-consistently determined profiles, the
valence level dips into the Dirac sea at  $M\approx750\,MeV$. This point is
shifted to $M\approx725$ for the reference profile. The deviation is
another evidence for the more sensitive dependence of valence and sea
contributions on details of the profile function, while their sum is quite
insensitive.  One should note, however, that the physically
relevant region for the constituent mass is around $M=400\,MeV$, where the
nucleon observables get reproduced by the reference profile very well.

The calculated nucleon observables are in sufficient agreement with
similar calculations \cite{Mei-91a,Mei-91,Wak-91}. The too small value of
the axial-vector coupling constant ($g_A\approx0.6\sim0.8)$ in comparison
to the experimental value ($g_A^{exp}\approx1.25$) is a lack shared by many
chiral models of the nucleon. However it is rather the aim of this paper
to compare between two theoretical approaches than to reproduce the
experimental values.

To complete our test of the reference profile we evaluate corrections to
the static hedgehog configurations due to zero-point modes and spin-isospin
quantization. Fig.~3 shows the energy $E_{CMM}$ (\ref{32}) of the
center-of-mass motion, the moment of inertia $I$ including the components
(\ref{35}) and (\ref{36}), and the resulting nucleon-delta mass splitting
(\ref{38}).  We establish an excellent agreement between the values for
both kinds of profiles in the physically relevant mass region below
600\,MeV.  Larger deviations appear at $M\gapp600\,MeV$. Here the energy
corrections are so large that the perturbative approach used for their
determination is already not justified.

\section{Conclusions}

We have self-consistently calculated average meson fields for the SU(2)
Nambu \& Jona-Lasinio model with scalar-isoscalar and
pseudoscalar-isovector couplings in Hartree approximation. The fields are
restricted to the chiral circle and to hedgehog configurations.  Infinite
quark contributions are  regularized within Schwinger's poper-time scheme.

The numerically determined meson profile functions turn out to be nearly
independent of the constituent quark mass. They can be approximated
quite well by a unique reference profile given by a simple arithmetic
expression, which interpolates between the correct asymptotic behaviour at
large and small radii.  It is shown that the reference profile does not
only approximate the self-consistent profiles but also reproduces the
relevant observables of the quark and meson configurations.

We conclude that many of the properties of the Nambu \& Jona-Lasinio
lagrangian can be studied using the reference profile instead of
applying the time-consuming determination of the self-consistent profile.
Changing the constituent quark mass $M$ only the strength $g$ of the
quark-meson coupling is influenced ($g\sim M$), while shape of the
meson fields is almost independent of $M$. The absolute strength of the
meson fields was fixed by the restriction to the chiral circle which can be
justified from an extended NJL model implementing the trace anomaly of QCD
\cite{Rip-92,Mei-93a,Wei-93}. If an accurate determination of the
self-consistent  profile is necessary, the reference profile may serve as a
suitable starting profile.

The reference profile can be compared with the  Woods-Saxon potential
describing the average field inside an atomic nucleus. Most of the nuclear
properties are sufficiently well described by this model potential which
is rather determined by a simple ansatz (with 3 parameters) than by a
sophisticated Hartree or Hartree-Fock procedure.\\[3mm]

\noindent{\small R.~W. thanks for repeated hospitality of the University of
Bochum.  The work has partially been supported by the Bundesministerium
f\"ur Forschung und Technologie, Bonn (contract 06 DR 107), the
COSY-Projekt of the KFA J\"ulich the Alexander-von-Humboldt-Foundation
(Feodeor-Lynen program) and by the Department of Energy (grant
DE-FG06-90ER40561).}

\newpage

\begin{center}{\large FIGURE CAPTIONS}\end{center}
\begin{enumerate}
\item
Self-consistet profiles in the mass region $350\,MeV\le M\le1000\,MeV$
All self-consistently calculated profiles fit in the area marked
by the {\em broken lines}.\\
The {\em full line} represents the reference profile (\ref{39}) with
$R=0.42\,fm$.

\item
Nucleon observables in dependence on the constituent quark mass $M$
calculated with self-consistently determined profiles ({\em full lines}) in
comparison to the reference profile $\Theta^{Ref}(R;r)$ defined in
eq.~(\ref{39}) with $R=0.42\,fm$.  ({\em broken lines})\\[2mm]
{\em Upper part}: Total energy $E$ (\ref{28}) and its components $E_{val}$
(\ref{29}), $E_{sea}^{Reg}$ (\ref{30}) and $E^{m}$ (\ref{31}).  \\
{\em Central part}: mean square radius $\bar{R}$ (\ref{23}) and its valence
($\bar{R}_{val}$) and sea contributions ($\bar{R}_{sea}$) calculated with
the corresponding densities (\ref{21}) and (\ref{22}), respectively.\\
{\em Lower part}: Proton axial-vector coupling constant $g_A$ (\ref{27})
(total value only).

\item
Energy corrections to the static hedgehog energy calculated with
self-consistently determined profiles ({\em full lines}) and
with the reference profile ({\em broken lines}) in dependence on the
constituent quark mass $M$.\\
{\em Upper part}: Energy of the center-of-mass motion
$E_{CMM}$ (\ref{32}).\\
{\em Central part}: Moment of inertia $I$ as the sum of the valence part
$I_{val}$ (\ref{35}) and the sea contributions $I_{sea}$ (\ref{36}).\\ {\em
Lower part}: Delta-nucleon mass splitting (\ref{38}).  \end{enumerate}

\newpage

\end{document}